\documentclass[aps,twocolumn,showpacs,superscriptaddress,groupedaddress,floatfix,
onecolumn,aps,prd,amsmath,amssymb]{revtex4}
\usepackage[english]{babel}
\usepackage[utf8]{inputenc}
\usepackage{epsfig}
\usepackage{multirow}
\usepackage{graphicx}
\usepackage{graphics}
\usepackage{subfigure}
\usepackage{epstopdf}
\usepackage{float}
\usepackage{bm}
\usepackage{braket}
\usepackage{lineno}
\usepackage{slashed}
\usepackage{hyperref}
\hypersetup{colorlinks=true, linkcolor=red, filecolor=black, citecolor=blue}

\begin{document}

\title{The two-gap BCS model in the large-$N$ approximation within a field-theory approach.}

\author{Leandro Nascimento}
\email{lon@ufpa.br}
\affiliation{Faculdade de Física, Universidade Federal do Pará, Avenida Augusto Correa 01, 66075-110, Belém, Pará,  Brazil}

\date{\today}

\begin{abstract}
We study the continuum version of the two-gap BCS model in (3+1)D within the large-N approximation. We calculate the effective potential of the model which depends on two independent energy gaps $\sigma$ and $\Delta$, where $\sigma$ describes the Cooper pair made of electrons that belong to the same internal symmetry whereas $\Delta$ describes the Cooper pair of electrons that have  a different internal symmetry. The effective potential is calculated by considering that the Debye frequency is an ultraviolet cutoff $\Lambda$, which is meant to describe the physical lattice of 3D superconductors. Our main result shows that the extra gap provides a possible inter-band phase transition that may be either a stable or metastable phase, depending on the competion between the coupling constants of the model. We also derive the critical temperature below which the phases may be observed.\end{abstract}

\maketitle

\section*{INTRODUCTION}

\textbf{Introduction}. - The theoretical description of quantum states of matter usually relies on the investigation of spontaneous symmetry breaking either in quantum field theory (QFT) \cite{Matthew,Coleman,Cahill} or condensed matter physics (CMP) \cite{livroMarino,Boulevard,Bernevig}. In both cases, it is the so-called effective potential that provides a gap equation, i.e, the equation for the order parameter. The BCS-superconductivity of 3D-semiconductors is one famous example of such description \cite{BCSoriginal} within CMP. The physical properties of a superconductor, within the BCS theory \cite{BCSoriginal}, are explained by a spontaneous Gauge-symmetry breaking due to the formation of Cooper pairs, i.e, the pairing of electrons with opposite momentum and spins. This occurs when mechanical vibrations become stronger than electronic repulsion at low temperatures, below a certain critical value $T_c$. Within this phase, magnetic fields are expelled by the material, which is known as Meissner-Ochsenfeld effect, and the application of a strong magnetic field leads the material to the normal phase. On the other hand, the theoretical explanation of high-temperature superconductivity remains an open question \cite{Boulevard,livroMarino,2GMarino}. In this context, new models and experiments have been proposed, in particular, for describing the contribution of an anisotropy in the electron-electron pairing, which is sometimes called a two-gap superconductor \cite{2GBCS, NJP,QM}. Recently,  a two-gap nonequilibrium superconductor system also has been observed due to the pairing of electrons that belong to different \textit{valleys} and \textit{bands} in a two-dimensional material \cite{Chamom}. This represents, from the point of view of symmetries, a breaking of the flavor symmetry of the matter field.

Here, we propose a field-theory approach to describe a flavor symmetry breaking in the BCS model, which implies a two-gap phase. Our approach, nevertheless, neglects the underlying mechanism for the flavor symmetry breaking, which is assumed to be generated by an external field, similar to the polarized light in the case of states out of equilibrium \cite{Chamom}. Thereafter, we use the Hubbard-Stratonovich transformation for obtaining a trilinear action with two auxiliary fields $\sigma$ and $\Delta$. Integrating out the fermions and considering a constant-field configuration, we derive the so-called effective potential of the model, which is dependent on these two fields. From this potential, we derive a set of two coupled-gap equations, which allows to calculate the vacuum expectation values of $\sigma$ and $\Delta$ as well as the critical temperature below each the phase transition may be observed. We show that the effective potential may have a stable and metastlable vacuum configuration, depending on the values of the coupling constants.


\textit{Content}: In this letter, we apply a field-theory approach to the standard BCS model in order to illustrate our large-N approximation. Thereafter, we apply this method to the large-$N$ approximation in a two-gap BCS model with internal symmetry breaking. This is a first step in order to describe a recent reported nonequilibrium superconductivity in lower-dimensional materials.



\textbf{The Continuum BCS Theory}. - In this section we apply the large-$N$ expansion \cite{Coleman} for the \textit{continuum} limit of the BCS model in order to derive the gap equation as well as the critical temperature. In order to do so, let us start with the action \cite{livroMarino}, given by
\begin{equation}
{\cal L}_{{\rm BCS}}=\psi^*_{s a}\left(i\frac{\partial}{\partial t}+\frac{\nabla^2}{2m}\right)\psi_{s a}-\frac{\lambda}{N} \psi^*_{\uparrow a}\psi^*_{\downarrow a}\psi_{\downarrow a}\psi_{\uparrow a}, \label{BCS}
\end{equation}
where $s=\uparrow,\downarrow$ describes the spin and $a=1,...,N$ is a flavor index. The matter field $\psi_{s a}$ describes the electronic excitation in a 3D-semiconductor with parabolic energy dispersion, namely, $\xi_{k}=\textbf{k}^2/2m$, where $m$ is the effective mass of the quasiparticle. The coupling constant $\lambda$ is meant to describe the electron-phonon interaction. We assume that there exist $N$ fields, such that we are allowed to use the large-$N$ expansion in Eq.~(\ref{BCS}). For the sake of simplicity, we shall consider $\hbar=1$.

It is well known that we can convert the quartic interaction in Eq.~(\ref{BCS}) into a trilinear one, using the Hubbard-Stratonovitch transformation with an auxiliary field $\sigma(x)$ \cite{Coleman}. Hence,
\begin{equation}
{\cal L}_{{\rm BCS}}\rightarrow{\cal L}_{{\rm BCS}}+\frac{N}{\lambda}\left(\sigma+\frac{\lambda}{N}\psi^*_{\uparrow a}\psi^*_{\downarrow a}\right)\left(\sigma^*+\frac{\lambda}{N}\psi_{\downarrow a}\psi_{\uparrow a}\right). \label{BCS2}
\end{equation}
On the other hand, the equation of motion for $\sigma$ is given by
\begin{equation}
\frac{\delta {\cal L}_{{\rm BCS}}}{\delta \sigma^*}|_{\sigma=\sigma_0}=0. \label{MEsig}
\end{equation}
Using Eq.~(\ref{BCS2}) in Eq.~(\ref{MEsig}), we find
\begin{equation}
\sigma_0=-\frac{\lambda}{N}\langle \psi^*_{\uparrow a}\psi^*_{\downarrow a} \rangle, \label{OP}
\end{equation}
which is our order parameter for the phase transition. Indeed, despite the fact that the action in Eq.~(\ref{BCS}) is invariant under the gauge transform $\psi_{sa}\rightarrow e^{i \alpha} \psi_{sa}$, the vacuum state breaks this symmetry when $\sigma_0\neq 0$, accordingly to Eq.~(\ref{OP}). In particular, for $\sigma_0\neq 0$, the system exhibits a pair of bounded electrons with opposite spins, the so-called Cooper pairs \cite{Boulevard}.

From Eq.~(\ref{BCS2}), we have our large-$N$ version of the continuum BCS theory, namely,
\begin{eqnarray}
{\cal L}_{{\rm BCS}}&=&\psi^*_{sa}\left(i\frac{\partial}{\partial t}+\frac{\nabla^2}{2m}\right)\psi_{sa} \nonumber\\
&+&\frac{N|\sigma|^2}{\lambda}+\sigma \psi_{\downarrow a}\psi_{\uparrow a}+\sigma^*\psi^*_{\uparrow a}\psi^*_{\downarrow a}. \label{BCSN2}
\end{eqnarray}
Eq.~(\ref{BCSN2}) is useful for calculating both the effective potential and the effective action in the large-$N$ limit, as we shall discuss.


\textbf{Effective Potential}. - Firstly, we note that Eq.~(\ref{BCSN2}) may be written as a quadratic action in terms of the Nambu field, namely, $\Phi^\dagger_a=(\psi^*_{\uparrow a} \,\, \psi_{\downarrow a})$. Hence, the partition function of the model reads
\begin{equation}
Z=\int D\Phi^\dagger_a D\Phi_a D\sigma e^{iS_{{\rm BCS}}[\Phi_a,\sigma]}. \label{ZBCS}
\end{equation}
The integration over $\Phi_a$ is solved with the help of the following property \cite{Boulevard}
\begin{equation}
\int D\Phi^\dagger_a D\Phi_a e^{-\int d^4 x \Phi^\dagger_a \hat{G} \Phi_a}=\exp\{N \ln \det \hat{G}\}, \label{ZBCS}
\end{equation}
where $\hat{G}$ is an arbitrary matrix. Therefore Eq.~(\ref{ZBCS}), yields
\begin{equation}
Z=\int D \sigma e^{i\int d^4 x {\cal L}_{{\rm eff}}[\sigma]},
\end{equation}
where the effective action for $\sigma$ is given by
\begin{equation}
{\cal L}_{{\rm eff}}[\sigma]=\frac{N|\sigma|^2}{\lambda}-i N\ln \det{\{-i\hat{K}[\sigma]\}}, \label{effsig}
\end{equation}
where $\hat{K}[\sigma]$ is a two-by-two matrix obtained from Eq.~(\ref{BCSN2}).
On the other hand, the effective potential $V_{{\rm eff}}[\sigma]$ is obtained through ${\cal L}_{{\rm eff}}[\sigma]$ when we consider a constant-field configuration, i.e, $\sigma(x)=\sigma$, where $\sigma$ is not dependent on space-time coordinates \cite{Boulevard}. Having this in mind, it follows that
\begin{equation}
Z=\int D \sigma e^{i\int d^4 x {\cal L}_{{\rm eff}}[\sigma]}\rightarrow \int D\sigma\, e^{iV_{{\rm eff}}[\sigma] \Omega}, \label{Veffdef}
\end{equation}
where $\Omega$ is an arbitrary space-time volume of quantization. Furtheremore, we have that $\det{\{-i\hat{K}[\sigma]\}}=-\omega^2+\xi_k^2+|\sigma|^2$. Hence, using Eq.~(\ref{effsig}) in Eq.~(\ref{Veffdef}), we find
\begin{eqnarray}
V_{{\rm eff}}[\sigma]&=&\frac{N |\sigma|^2}{\lambda} \nonumber \\
&-&i N \int\frac{d^4 k}{(2\pi)^4}\ln(-\omega^2+\xi_k^2+|\sigma|^2), \label{Veff1}
\end{eqnarray}
where $d^4k=d\omega d^3 \textbf{k}$ and $\xi_k=\textbf{k}^2/2m$. Note that the last term in the rhs of Eq.~(\ref{Veff1}) is the Fourier transform of $\ln \det{\{-i\hat{K}[\sigma]\}}$ for $\sigma(x)=\sigma$. From Eq.~(\ref{Veff1}), we conclude that the effective potential is invariant under the continuous gauge transformation, therefore, the vacuum state is expected to be infinitely degenerated. Consequently, the effective potential is invariant also under a restricted transformation $\sigma\rightarrow-\sigma$. This is the same feature shown by a generic Landau-Ginzburg potential \cite{Boulevard}, as we shall prove later.

\textbf{The Gap Equation}. - The gap equation is obtained by extremizing the effective potential, i.e.,
\begin{equation}
\frac{\partial V_{{\rm eff}}}{\partial \sigma}|_{\sigma=\sigma_0}=V'[\sigma_0]=0.
\end{equation}
Using Eq.~(\ref{Veff1}), we find
\begin{eqnarray}
-\frac{1}{\lambda}=i \int\frac{d^4 k}{(2\pi)^4}\frac{1}{\omega^2-\xi_k^2-|\sigma_0|^2}. \label{BCSgap01}
\end{eqnarray}
After integrating over $\omega$, using $\omega^2\rightarrow \omega^2-i\delta$ with $\delta\rightarrow 0^+$ in Eq.~(\ref{BCSgap01}), we obtain
\begin{eqnarray}
-\frac{1}{\lambda}=-\frac{1}{2} \int\frac{d^3 k}{(2\pi)^3}\frac{1}{\sqrt{\xi_k^2+|\sigma_0|^2}}, \label{BCSgapk}
\end{eqnarray}
which is the so-called BCS gap equation \cite{Boulevard}. Obviously, Eq.~(\ref{BCSgapk}) is divergent for $k\rightarrow \infty$, as it usually is the case for gap equations. We shall circumvent this problem by considering an ultraviolet cutoff $\Lambda$.

Note that the integral variable may be written as
\begin{equation}
\frac{d^3 k}{(2\pi)^3}=N_s D(\xi_k) d\xi_k,
\end{equation}
where 
\begin{equation}
D(\xi_k)=\frac{(2m)^{3/2}\xi_k^{1/2}}{4\pi^2}
\end{equation}
is the density of states, $\xi_k=\textbf{k}^2/2m$, and $N_s=2$ is the spin degeneracy. The standard approximation for the BCS model is to consider that $D(\xi_k)\approx D(k_F)=m k_F/ 2\pi^2$, where $k_F$ is the Fermi momentum. $D(k_F)$ is, therefore, the density of states at the Fermi surface of the metallic state.  It turns out that for a metal $k_F \approx 10^4$K in units of temperature. On the other hand, phonons may only transfer energy close to the Debye frequency $ \omega_D \propto 1/a \approx 10^2$K, where $a$ is the lattice parameter. Therefore, we conclude that the integral may be calculated in the interval $\xi_k \in [0,\omega_D]\equiv [0,\Lambda]$, where $\Lambda=\omega_D$ is an ultraviolet cutoff. We conclude, based on these numbers, that most of the electrons are not affected by this interaction, except those who are close to the Fermi energy. This explains our approximation into the integral over $k$, hence, it follows that \cite{Boulevard,livroMarino} 
\begin{equation}
\int \frac{d^3k}{(2\pi)^3}\rightarrow \int_{0}^{\omega_D} d\xi_k N_s D(k_F). \label{app1}
\end{equation}
Using Eq.~(\ref{app1}) in Eq.~(\ref{BCSgapk}), we find
\begin{eqnarray}
\frac{1}{\lambda}\approx \int_{0}^{\Lambda} d\xi_k \frac{D(k_F)}{\sqrt{\xi_k^2+|\sigma_0|^2}}=D(k_F)\sinh^{-1}\left[\frac{\Lambda}{|\sigma_0|}\right]. \label{BCSgapxik}
\end{eqnarray}
Finally, because $\sinh^{-1}[\Lambda/|\sigma_0|]\approx \ln[2\Lambda/|\sigma_0|]$ for $\Lambda \gg |\sigma_0|$, we find
\begin{equation}
|\sigma_0|\approx 2 \Lambda e^{-1/\lambda D(k_F)}, \label{BCSgap}
\end{equation}
which is the BCS gap for a superconductor. 

Note that, by comparison between Eq.~(\ref{BCSgap01}) and Eq.~(\ref{BCSgapxik}), we may summarize all of our approximations by using
\begin{eqnarray}
&i&\int \frac{d^4 k}{(2\pi)^4}\frac{1}{\omega^2-\xi_k^2-|Z|^2}\approx   \nonumber \\
&-&D(k_F) \ln\left(\frac{2\Lambda}{|Z| }\right)\label{appfull}
\end{eqnarray}
for any $Z$ and $\Lambda \gg |Z|$. 
Eq.~(\ref{appfull}) is very useful for our purposes. Indeed, it allow us to easily find an analytical expression for $V_{{\rm eff}}[\sigma]$ in Eq.~(\ref{Veff1}). 

\begin{figure}[htb]
\centering
\includegraphics[scale=0.8]{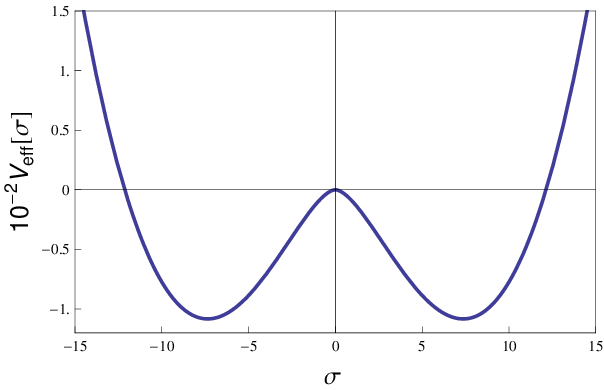}
\caption{(Color online) \textbf{The Effective Potential  $V_{{\rm eff}}[\sigma]$}. We plot the function in  Eq.~(\ref{VeffBCSend}) with $\Lambda=10$ (units of energy), $\lambda=1.0$ (units of $\Lambda^{-2}$), $D(k_F)=1$ and $N=4$. It is shown that the two solutions $\pm |\sigma_0|$, given by Eq.~(\ref{BCSgap}), of the gap equation $V'[\sigma_0]=0$ are energetically favorable in comparison with the symmetric solution $\sigma=0$. This resembles the well known Landau-Ginzburg potental.} \label{fig1}
\end{figure}

We assume that $\sigma$ is real, such that $|\sigma|^2=\sigma^2$. Next, we calculate the derivative of Eq.~(\ref{Veff1}) in respect to $\sigma$ to find
\begin{eqnarray}
\frac{\partial V_{{\rm eff}}}{\partial \sigma}&=&\frac{2N \sigma}{\lambda} \nonumber \\
&+&i N\int \frac{d^4 k}{(2\pi)^4}\frac{2 \sigma}{\omega^2-\xi_k^2-|\sigma|^2}. \label{DVeff1}
\end{eqnarray}
Using Eq.~(\ref{appfull}) in Eq.~(\ref{DVeff1}), we have
\begin{eqnarray}
\frac{\partial V_{{\rm eff}}}{\partial \sigma}&=&\frac{2N \sigma}{\lambda} \nonumber \\
&-& 2N D(k_F) \sigma \ln\left(\frac{2\Lambda}{|\sigma| }\right). \label{DVeff2}
\end{eqnarray}
In order to calculate $V_{{\rm eff}}[\sigma]$, we use the identity
\begin{equation}
V_{{\rm eff}}[\sigma]=\int_0^\sigma \frac{\partial V_{{\rm eff}}}{\partial \sigma'} d\sigma'+V_0, \label{idenV}
\end{equation}
where $V_0$ is an arbitrary constant, which we consider to vanish, without loss of generality. Finally, using Eq.~(\ref{DVeff2}) in Eq.~(\ref{idenV}) and integrating over $\sigma'$, we obtain
\begin{equation}
V_{{\rm eff}}[\sigma]=\frac{N \sigma^2}{\lambda}-\frac{N D(k_F)\sigma^2}{2}-N D(k_F) \sigma^2 \ln\left(\frac{2\Lambda}{|\sigma|}\right). \label{VeffBCSend}
\end{equation}
In Fig.~\ref{fig1}, we plot the effective potential and discuss the energetically favorable solution. The main results are very close to the Landau-Ginzburg theory.

\textbf{The Critical Temperature}. - Here, we derive the critical temperature of the BCS model. In order to do so, we return to its effective potential at zero temperature in Eq.~(\ref{Veff1}). Nevertheless, here, we separate the $\omega$ and $k$ integrals, hence,
\begin{equation}
V_{{\rm eff}}=\frac{N|\sigma|^2}{\lambda}- i N \int \frac{d^3k}{(2\pi)^3} \int \frac{d\omega}{2\pi} \ln(-\omega^2+\xi_k^2+|\sigma|^2). \label{VeffBCS2}
\end{equation}
We shall introduce the Matsubara frequencies in the so-called imaginary-time formalism. First, let us perform a Wick rotation given by $\omega\rightarrow -i\omega$. Thereafter, we use $\omega\rightarrow \omega_n=(2n+1)\pi T$, where $n=0,1,2,..$ is an integer and
\begin{equation}
\int \frac{d\omega}{2\pi} (...)\rightarrow T \sum_n (...).
\end{equation}
Therefore, Eq.~(\ref{VeffBCS2}) reads
\begin{equation}
V_{{\rm eff}}[\sigma,T]=\frac{N|\sigma|^2}{\lambda}-N T \sum^{+\infty}_{n=-\infty}\int \frac{d^3k}{(2\pi)^3}  \ln(\omega_n^2+\xi_k^2+|\sigma|^2), \label{VeffBCST}
\end{equation}
which is the effective potential of the BCS model at finite temperature.

From the first derivative of Eq.~(\ref{VeffBCST}) in respect to $\sigma$ calculated at $\sigma=\sigma_0$, we find
\begin{equation}
\frac{1}{\lambda}= T \sum^{+\infty}_{n=-\infty}\int \frac{d^3k}{(2\pi)^3} \frac{1}{\omega_n^2+\xi_k^2+|\sigma_0|^2}. \label{GapBCST}
\end{equation}
Next, we use the following identity
\begin{equation}
\sum^{+\infty}_{n=-\infty}  \frac{1}{\omega_n^2+\xi_k^2+|\sigma_0|^2}=\frac{\tanh\left(\sqrt{\xi_k^2+\sigma_0^2}/2 T\right)}{2 T \sqrt{\xi_k^2+\sigma_0^2}} \label{prop1}
\end{equation}
in order to solve the Matsubara sum in Eq.~(\ref{GapBCST}). On the other hand, we may use that $\tanh(\beta z/2)=1-2n_F(z)$ for $z\equiv \sqrt{\xi_k^2+\sigma_0^2}$, $\beta=1/T$, where $n_F(z)=1/(1+e^{\beta z})$ is the Fermi-Dirac distribution in Eq.~(\ref{prop1}). Having these properties in mind, Eq.~(\ref{GapBCST}) yields
\begin{eqnarray}
\frac{1}{\lambda}&=&\frac{1}{2} \int\frac{d^3 k}{(2\pi)^3}\frac{1}{\sqrt{\xi_k^2+|\sigma_0|^2}}\nonumber\\
&-&\int\frac{d^3 k}{(2\pi)^3}\frac{n_F(\sqrt{\xi_k^2+\sigma_0^2})}{\sqrt{\xi_k^2+\sigma_0^2}}. \label{GapBCST2}
\end{eqnarray}
Clearly, the first term in the rhs of Eq.~(\ref{GapBCST2}) is the gap equation at zero temperature given by Eq.~(\ref{BCSgapk}). Furthermore, the second term describes the effects of the thermal bath and goes to zero as $T\rightarrow 0$.

Similarly to the previous case, we apply the approximation $\int \frac{d^3 k}{(2\pi)^3}\rightarrow  \int_{0}^{\Lambda} d\xi_k N_s D(k_F)$. Therefore, Eq.~(\ref{GapBCST}) reads
\begin{equation}
\frac{1}{\lambda}\approx D(k_F) \int_{0}^{\Lambda} d\xi_k \frac{\tanh\left(\sqrt{\xi_k^2+\sigma_0^2}/2 T\right)}{\sqrt{\xi_k^2+\sigma_0^2}}, \label{GapTc}
\end{equation}
where we have replaced $\omega_D\rightarrow \Lambda$. The critical temperature $T_c$ is defined as the temperature in which the gap $\sigma_0$ vanishes. This is determined by Eq.~(\ref{GapTc}), hence,
\begin{equation}
\frac{1}{\lambda}\approx D(k_F) \int_{0}^{\Lambda} d\xi_k \frac{\tanh\left(\xi_k/2 T_c\right)}{\xi_k}.\label{Tc}
\end{equation}

Next, let us define $y\equiv \xi_k/(2 T_c)$. Therefore, Eq.~(\ref{Tc}) reads
\begin{equation}
\frac{1}{\lambda}\approx D(k_F) \int_{0}^{\Lambda/(2T_c)} dy \frac{\tanh(y)}{y}. \label{Tc2}
\end{equation}
The integral over $y$ may be calculated by parts, i.e,,
\begin{eqnarray}
\int_{0}^{\Lambda/(2T_c)} dy \frac{\tanh(y)}{y}&=&\tanh(y)\ln(y)|{_0^{\Lambda/2T_c}} \nonumber \\
&-&\int_{0}^{\Lambda/(2T_c)} dy \frac{\ln(y)}{\cosh^2(y)}. \label{inty}
\end{eqnarray}
Within the regime $\Lambda \gg 2T_c$, Eq.~(\ref{inty}) yields
\begin{eqnarray}
\int_{0}^{\Lambda/(2T_c)} dy \frac{\tanh(y)}{y}&\approx&\ln\left(\frac{\Lambda}{2T_c}\right)-\ln\left(\frac{\pi}{4e^{\gamma_E}}\right) \nonumber \\
&=& \ln\left(\frac{2\Lambda e^{\gamma_E} }{ \pi T_c}\right), \label{inty2}
\end{eqnarray}
where $\gamma_E\approx 0.57$ is the Euler constant. Using Eq.~(\ref{inty2}) in Eq.~(\ref{Tc}), we find
\begin{equation}
T_c\approx \frac{2\Lambda e^{\gamma_E}}{\pi} e^{-1/\lambda D(k_F)}=|\sigma_0| \frac{e^{\gamma_E}}{\pi}, \label{Tcend}
\end{equation}
where in the last term in the rhs of Eq.~(\ref{Tcend}) we have used the expression for the gap in Eq.~(\ref{BCSgap}). This shows that an universal ratio for BCS superconductors is found, namely, $|\sigma_0|/(k_B T_c)= \pi/e^{\gamma_E}\approx 1.76$, where we have properly included the Stefan-Boltzmann constant $k_B$. Finally, it is well known that the typical values of $T_c$ are on the interval 30-40K, which can not describe the high-$T_c$ superconductors where $T_c$ may be close to 100K \cite{Boulevard, livroMarino}.

\textbf{The Two-Gap BCS Model.} - In this section we propose a generalization of the BCS model, considering the introduction of an internal symmetry breaking. Thereafter, we study the formation of Cooper pairs. This model describes a two-gap superconductor made of two nonrelativistic electrons, similarly to what has been done in Ref.~\cite{2GBCS} for underdoped cuprate superconductors. 

Our first step is to assume that the matter field is described by three indexes, i.e., $\psi=\psi_{ais}$, where $a=1,...,N$ is the flavor index, $i=K,K'$ describes an internal symmetry of the cristal, such as a sublattice symmetry, and $s=\uparrow,\downarrow$ describes de spin. Therefore, the total symmetry is described by matrices in the group SU(2)$\times$SU(N)$\times$SU(2). Hence, the two-gap-BCS Lagrangian reads  
\begin{eqnarray}
{\cal L}_{{\rm 2G}}&=&\psi^*_{ais}\left(i\frac{\partial}{\partial t}+\frac{\nabla^2}{2m}\right)\psi_{ais} \nonumber \\
&-&\frac{\lambda}{N} \psi^*_{a,K,\uparrow}\psi^*_{a,K,\downarrow}\psi_{a,K,\downarrow}\psi_{a,K,\uparrow} \nonumber \\
&-& \frac{\lambda}{N} \psi^*_{a,K',\uparrow}\psi^*_{a,K',\downarrow}\psi_{a,K',\downarrow}\psi_{a,K',\uparrow}\nonumber \\
&-&\frac{g}{N} \psi^*_{a,K,\uparrow}\psi^*_{a,K',\downarrow}\psi_{a,K',\downarrow}\psi_{a,K,\uparrow} \nonumber \\
&-&\frac{g}{N} \psi^*_{a,K',\uparrow}\psi^*_{a,K,\downarrow}\psi_{a,K,\downarrow}\psi_{a,K',\uparrow}, \label{2GBCS}
\end{eqnarray}
where we shall call $\lambda$ as intra-band coupling constant and $g$ as inter-band coupling constant in reference to the new indexes $K,K'$.

Similarly to what we did before, we apply the Hubbard-Stratonovich transform, namely,
\begin{eqnarray}
{\cal L}_{{\rm 2G}}\rightarrow{\cal L}_{{\rm 2G}}&+&\frac{N}{\lambda}\left(\sigma_K+\frac{\lambda}{N}\psi^*_{a,K,\uparrow }\psi^*_{a,K,\downarrow}\right) \nonumber\\
&\times &\left(\sigma^*_K+\frac{\lambda}{N}\psi_{a,K,\downarrow}\psi_{a,K,\uparrow}\right) \nonumber \\
&+&\frac{N}{\lambda}\left(\sigma_{K'}+\frac{\lambda}{N}\psi^*_{a,K',\uparrow }\psi^*_{a,K',\downarrow}\right) \nonumber\\
&\times &\left(\sigma^*_{K'}+\frac{\lambda}{N}\psi_{a,K',\downarrow}\psi_{a,K',\uparrow}\right) \nonumber \\ &+&\frac{N}{g}\left(\Delta_{KK'}+\frac{g}{N}\psi^*_{a,K,\uparrow }\psi^*_{a,K',\downarrow}\right) \nonumber\\
&\times &\left(\Delta^*_{KK'}+\frac{g}{N}\psi_{a,K',\downarrow}\psi_{a,K,\uparrow}\right) \nonumber \\
&+&\frac{N}{g}\left(\Delta_{K'K}+\frac{g}{N}\psi^*_{a,K',\uparrow }\psi^*_{a,K,\downarrow}\right) \nonumber\\
&\times &\left(\Delta^*_{K'K}+\frac{g}{N}\psi_{a,K,\downarrow}\psi_{a,K',\uparrow}\right). \label{HB2G}
\end{eqnarray}
From the equation of motion for the auxiliary fields $\sigma_K$, $\sigma_{K'}$, $\Delta_{KK'}$, and $\Delta_{K'K}$, we find
\begin{equation}
\sigma^0_K=-\frac{\lambda}{N} \langle \psi^*_{a,K,\downarrow}\psi^*_{a,K,\uparrow} \rangle,
\end{equation}
\begin{equation}
\sigma^0_{K'}=-\frac{\lambda}{N} \langle \psi^*_{a,K',\downarrow}\psi^*_{a,K',\uparrow} \rangle,
\end{equation}
\begin{equation}
\Delta^0_{KK'}=-\frac{g}{N} \langle \psi^*_{a,K',\downarrow}\psi^*_{a,K,\uparrow} \rangle,
\end{equation}
and
\begin{equation}
\Delta^0_{K'K}=-\frac{g}{N} \langle \psi^*_{a,K,\downarrow}\psi^*_{a,K',\uparrow} \rangle,
\end{equation}
which are our four order parameters for the two-gap BCS model, where the index $0$ refers to the vacuum expectation value for each of them. Note that the expectation values of $\sigma^0_K$ describes the intra-band symmetry breaking while $\Delta^0_{KK'}$ defines the inter-band symmetry breaking phase.

Following the same steps as before, after we use the Hubbard-Stratonovich transformation in Eq.~(\ref{HB2G}), we find
\begin{eqnarray}
{\cal L}_{{\rm 2G}}&=&\psi^*_{ais}\left(i\frac{\partial}{\partial t}+\frac{\nabla^2}{2m}\right)\psi_{ais}\nonumber \\
&+&\frac{N|\sigma_K|^2}{\lambda}+\frac{N |\sigma_{K'}|^2}{\lambda}+\frac{N |\Delta_{KK'}|^2}{g}+\frac{N|\Delta_{K'K}|^2}{g} \nonumber\\
&+&\sigma_K \psi_{a,K,\downarrow}\psi_{a,K,\uparrow}+\sigma_K^*\psi^*_{a,K,\uparrow}\psi^*_{a,K,\downarrow} \nonumber \\
&+&\sigma_{K'} \psi_{a,K',\downarrow}\psi_{a,K',\uparrow}+\sigma_{K'}^*\psi^*_{a,K',\uparrow}\psi^*_{a,K',\downarrow} \nonumber \\
&+& \Delta_{KK'} \psi_{a,K',\downarrow}\psi_{a,K,\uparrow}+\Delta^*_{KK'}\psi^*_{a,K,\uparrow}\psi^*_{a,K',\downarrow}\nonumber \\
&+&\Delta_{K'K} \psi_{a,K,\downarrow}\psi_{a,K',\uparrow}+\Delta^*_{K'K}\psi^*_{a,K',\uparrow}\psi^*_{a,K,\downarrow}. \label{2GBCS1}
\end{eqnarray}
Note that Eq.~(\ref{2GBCS1}) is quadratic in the Nambu field, namely, $\Phi^\dagger_a=(\psi^*_{a,K,\uparrow} \,\, \psi_{a,K,\downarrow} \,\, \psi^*_{a,K',\uparrow} \,\, \psi_{a,K',\downarrow})$. 


\textbf{The Effective Potential.} - The calculation follows the same steps as in the BCS theory. Therefore, after integrating over the Nambu field $\Phi^\dagger_a$, we have
\begin{eqnarray}
V_{{\rm eff}}^{2G}&=&\frac{2 N\sigma^2}{\lambda}+\frac{2 N\Delta^2}{g} \nonumber \\
&-& \frac{N D(k_F)X^2}{2}-N D(k_F) X^2 \ln\left(\frac{2\Lambda}{|X|}\right) \nonumber \\
&-& \frac{N D(k_F)Y^2}{2}-N D(k_F) Y^2 \ln\left(\frac{2\Lambda}{|Y|}\right), \label{Veff2G2}
\end{eqnarray}
where $X=\sigma+\Delta$ and $Y=\sigma-\Delta$.


\textbf{The Coupled-Gap Equations.} - The gap equation is now obtained by extremizing the effective potential in respect to both fields $\sigma$ and $\Delta$, i.e., we have to calculate
\begin{equation}
\frac{\partial V_{{\rm eff}}^{2G}}{\partial \sigma}|_{\sigma=\sigma_0,\Delta_0}=0 \label{gaps2}
\end{equation}
and
\begin{equation}
\frac{\partial V_{{\rm eff}}^{2G}}{\partial \Delta}|_{\Delta=\Delta_0,\sigma_0}=0. \label{gap2D}
\end{equation}
Clearly, Eq.~(\ref{gaps2}) and Eq.~(\ref{gap2D}) are two independent equations that provide the solutions for $\sigma_0$ and $\Delta_0$. These are the minimum of the effective potential. After summing Eq.~(\ref{gaps2}) and Eq.~(\ref{gap2D}), we find
\begin{equation}
4N\left[\frac{\Delta_0}{g}+\frac{\sigma_0}{\lambda}-D(k_F) X_0 \ln\left(\frac{2\Lambda}{|X_0|}\right)\right]=0, \label{gap2Gs}
\end{equation}
where $X_0=\sigma_0+\Delta_0$. On the other hand, by calculating the subtraction between Eq.~(\ref{gaps2}) and Eq.~(\ref{gap2D}), we obtain
\begin{equation}
4N\left[-\frac{\Delta_0}{g}+\frac{\sigma_0}{\lambda}-D(k_F) Y_0 \ln\left(\frac{2\Lambda}{|Y_0|}\right)\right]=0, \label{gap2Gd}
\end{equation}
where $Y_0=\sigma_0-\Delta_0$.

\textbf{The Vacuum Stability.}- The solutions of the coupled-gap equations in Eq.~(\ref{gap2Gs}) and Eq.~(\ref{gap2Gd}), with the mininum of effective potential in Eq.~(\ref{Veff2G2}), allow us to discuss the vacuum stability of the model in terms of the coupling constants $(\lambda,g)$. Firstly, let us consider the case when $\lambda=g$. Here, the effective potential is symmetric in its two variables, i.e, $V_{{\rm eff}}^{2G}[\sigma,\Delta]=V_{{\rm eff}}^{2G}[\Delta,\sigma]$, hence, the values of the energy gaps are expected to be equal at the ground state. This is shown in Fig.~2, where we plot the effective potential in terms of the two gaps. Note that the global mininums are located at $(\sigma,\Delta)=(\sigma_0,0)$ and $(\sigma,\Delta)=(0,\Delta_0)$, where $|\sigma_0|=|\Delta_0|=2\Lambda e^{-1/\lambda D(k_F)}$ is the stantard BCS gap. We conclude, therefore, that the model allows both stable vacuums, either an intra-band phase or an inter-band phases. 
\begin{figure}[htb]
\centering
\includegraphics[scale=0.7]{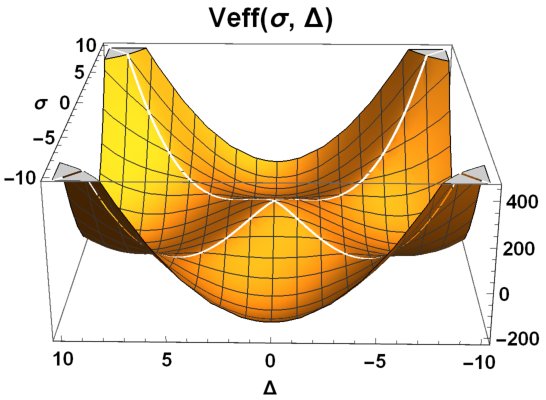}
\caption{(Color online) \textbf{The Effective Potential  $V_{{\rm eff}}^{2G}[\sigma,\Delta]$}. We plot the function in  Eq.~(\ref{Veff2G2}) with $\Lambda=10$ (units of energy), $\lambda=1.0$ (units of $\Lambda^{-2}$), $D(k_F)=1.0$, $g=1.0$ (units of $\Lambda^{-2}$), and $N=4$.} \label{fig2}
\end{figure}

Next, we discuss the case when $\lambda<g$, which provides a richer vacuum stability. From Fig.~3, we find two possible phases, namely, a local minimum at $(\sigma,\Delta)=(\sigma_0,0)$, where $|\sigma_0|=2\Lambda e^{-1/\lambda D(k_F)}$, and a global minimum at $(\sigma,\Delta)=(0,\Delta_0)$, where $|\Delta_0|=2\Lambda e^{-1/g D(k_F)}$. Note that as we increase $g$ in comparison to $\lambda$, the bottom along the line $\sigma=0$ becomes more deeper and, therefore, it is the true vacuum (global minimum) of the system. Furthermore, whenever $\lambda<g$, we have $V_{{\rm eff}}^{2G}[0,\Delta_0]<V_{{\rm eff}}^{2G}[\sigma_0,0]$, which means the system favors an inter-band phase. We also may consider the case when $\lambda>g$, which yields similar conclusions, but now the true vacuum is an intra-band phase.

The metastable (local minimum) phase along the line $\Delta=0$ is also reachable, providing an intra-band phase with $\sigma_0\neq 0$. We believe that such phenomenological model could be a relevant step for describing nonequilibrium superconductivity, where we can realize the pairing of electrons that belong to different symmetry indexes and control the strength of the coupling constants. 
\begin{figure}[htb]
\centering
\includegraphics[scale=0.73]{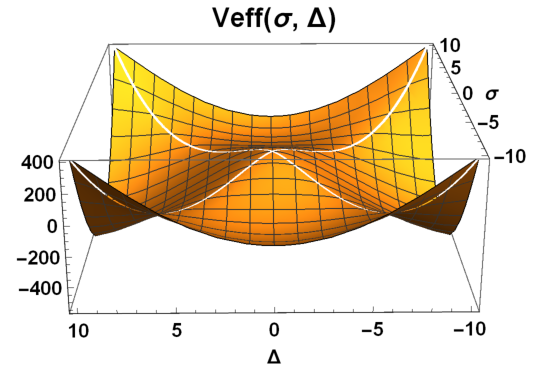}
\caption{(Color online) \textbf{The Effective Potential  $V_{{\rm eff}}^{2G}[\sigma,\Delta]$}. We plot the function in  Eq.~(\ref{Veff2G2}) with $\Lambda=10$ (units of energy), $\lambda=1.0$ (units of $\Lambda^{-2}$), $D(k_F)=1.0$, $g=2.0$ (units of $\Lambda^{-2}$), and $N=4$.} \label{fig2}
\end{figure}



\textbf{The Critical Temperature.} - We would like to calculate the critical temperatures for both intra- and inter-band phase transitions. After introducing the Matsubara frequencies, as we did before, we find 
\begin{eqnarray}
& & V_{{\rm eff}}^{2G}[\sigma,\Delta,T]=\frac{2 N|\sigma|^2}{\lambda}+\frac{2 N|\Delta|^2}{g} \nonumber \\
&-&NT \sum_{n=-\infty}^{\infty}\int\frac{d^3 k}{(2\pi)^3}\ln(\omega_n^2+\xi_k^2+|X|^2) \nonumber \\
&-&NT \sum_{n=-\infty}^{\infty}\int\frac{d^3 k}{(2\pi)^3}\ln(\omega_n^2+\xi_k^2+|Y|^2), \label{Veff2GT}
\end{eqnarray}
which is the effective potential of the two-gap BCS model at finite temperature. Similarly to the zero temperature case, we have two-coupled gap equations, namely,
\begin{equation}
\frac{\partial V_{{\rm eff}}^{2G}[T]}{\partial \sigma}|_{\sigma=\sigma_0,\Delta_0}=0 \label{gaps2T}
\end{equation}
and
\begin{equation}
\frac{\partial V_{{\rm eff}}^{2G}[T]}{\partial \Delta}|_{\Delta=\Delta_0,\sigma_0}=0. \label{gap2DT}
\end{equation}
For the sake of simplicity we assume that both $\sigma$ and $\Delta$ are real constants. Therefore, these gap equations are given by
\begin{eqnarray}
& &\frac{4 N \sigma_0}{\lambda}-NT \sum_{n=-\infty}^{\infty}\int\frac{d^3 k}{(2\pi)^3} \frac{2X_0}{\omega_n^2+\xi_k^2+|X_0|^2} \nonumber \\
&-&NT \sum_{n=-\infty}^{\infty}\int\frac{d^3 k}{(2\pi)^3} \frac{2|Y_0|}{\omega_n^2+\xi_k^2+|Y_0|^2}=0 \label{GapsT}
\end{eqnarray}
and
\begin{eqnarray}
& &\frac{4 N \Delta_0}{g}-NT \sum_{n=-\infty}^{\infty}\int\frac{d^3 k}{(2\pi)^3} \frac{2X_0}{\omega_n^2+\xi_k^2+|X_0|^2} \nonumber \\
&+&NT \sum_{n=-\infty}^{\infty}\int\frac{d^3 k}{(2\pi)^3} \frac{2|Y_0|}{\omega_n^2+\xi_k^2+|Y_0|^2}=0, \label{GapDT}
\end{eqnarray}
respectively.

From Eq.~(\ref{GapsT}) and Eq.~(\ref{GapDT}), we conclude that
\begin{eqnarray}
& &\frac{4 N \sigma_0}{\lambda}+\frac{4 N \Delta_0}{g} \nonumber \\
&-&2NT \sum_{n=-\infty}^{\infty}\int\frac{d^3 k}{(2\pi)^3} \frac{2X_0}{\omega_n^2+\xi_k^2+|X_0|^2}=0.\label{GapTend}
\end{eqnarray}
The sum over $\omega_n$ may be solved with the help of Eq.~(\ref{prop1}). Furthermore, we also replace $\int \frac{d^3 k}{(2\pi)^3}\rightarrow  \int_{0}^{\Lambda} d\xi_k N_s D(k_F)$ to obtain
\begin{eqnarray}
& &\frac{\sigma_0}{\lambda}+\frac{ \Delta_0}{g} \nonumber \\
&\approx & X_0 D(k_F) \int_{0}^\Lambda  d\xi_k \frac{\tanh\left[\sqrt{\xi^2_k+X^2_0}/(2T)\right]}{\sqrt{\xi^2_k+X^2_0}}. \label{GapTend2}
\end{eqnarray}
Eq.~(\ref{GapTend2}) provides both $T_{c1}$ and $T_{c2}$, the critical temperature for both the intra-band and inter-band phase transition, respectively.

Accordingly to the results of the previous section, the intra-band phase occurs at $\Delta_0=0$. On the other hand, we definie $T_{c1}$ as the temperature in which $\sigma_0$ vanishes, hence, using these assumptions in Eq.~(\ref{GapTend2}) it follows that
\begin{equation}
T_{c1}\approx \frac{2\Lambda e^{\gamma_E}}{\pi} e^{-1/\lambda D(k_F)}=\sigma_0 \frac{e^{\gamma_E}}{\pi}, \label{Tc1end}
\end{equation}
where we have considered the same set of approximations as we have made in the standard BCS model. Similarly, we also may obtain that
\begin{equation}
T_{c2}\approx \frac{2\Lambda e^{\gamma_E}}{\pi} e^{-1/gD(k_F)}=\Delta_0 \frac{e^{\gamma_E}}{\pi}. \label{Tc2end}
\end{equation}
As expected, when $\lambda=g$, we have $T_{c1}=T_{c2}$ and the model admits both intra- and inter-band phase transitions when $T<T_{c1}=T_{c2}$. For $\lambda<g$, we have $T_{c1}<T_{c2}$, accordingly to Eq.~(\ref{Tc1end}) and Eq.~(\ref{Tc2end}). In this case, for $T<T_{c1}<T_{c2}$, the system admits either the intra- and inter-band phases, as discussed earlier. For $T_{c1}<T<T_{c2}$, we only find the inter-band phasse while for $T>T_{c2}$ we have the normal phase.



\textbf{Conclusions and Discussions.} - In this work we have described the solutions of a two-gap problem in a continuum version of the BCS model, using a finite ultraviolet cutoff $\Lambda$. Our main result shows that the extra gap provides a possible inter-band phase transition that may be either a stable or metastable phase depending on the coupling constants. We also derive the critical temperature in which each phase may be observed. These conclusions are corroborated by an analysis of the effective potential at the large-$N$ approximation. 
The two-gap BCS model \cite{2GBCS} is expected to be relevant for describing unconventional pairing in superconductors with high-$T_c$, for example, in MgB2 and underdoped cuprates \cite{MgB2}. This, unfortunately, has been less discussed in literature. Here, we gave a step forward by considering its continuum version.

The large-$N$ approximation that we applied in this work resembles more quantum-field-theory methods than condensed matter physics.
This happens because we have neglected several microscopic informations about hybridization, for example, of the electronic bands. In principle, one concludes that this level of approximation is too far from reality. Fortunately, this seems not to be the case. Indeed, it has been shown that a quantum-field-theory description of the Fermi velocity renormalization in graphene \cite{Voz} yields a good agreement with experimental data \cite{Elias}. Furthermore, the application of Pseudo quantum electrodynamics \cite{BJP} for describing excitonic spectrum (pairs of electron and hole) in transition metal dichalcogenides has also been shown very useful \cite{Exc}. Therefore, it is clear that there exist a window of applicability for such approximations in electronic properties of a crystal. In general, for a better comparison with experiments, we may assume that we must neglect impurities, disorder, lattice vibrations, and higher-order-momentum dependence in the energy dispersion. Although it is possible to improve the approximations for taking into account all of these effects, a full model describing these features is yet to be found.









\acknowledgments
L. O. N. is partially supported by Conselho Nacional de Desenvolvimento Cientifico e Tecnologico (CNPq) and by CAPES/NUFFIC, finance code 0112. L. O. N. thanks V. S. Alves, C. M. Smith, and E. C. Marino for several insightful discussions.

\end{document}